\documentstyle[preprint,aps]{revtex}

\begin{document}

\draft

\preprint{UTPT-94-36}

\title{Field Equations and Conservation Laws in the Nonsymmetric
Gravitational Theory}

\author{J. L\'egar\'e and J. W. Moffat}

\address{Department of Physics, University of Toronto,
Toronto, Ontario, Canada M5S 1A7}

\date{December 7, 1994}

\maketitle

\begin{abstract}

The field equations in the nonsymmetric gravitational theory
are derived from a Lagrangian density using a first-order formalism.
Using the general covariance of the Lagrangian density, conservation
laws and tensor identities
are derived.
Among these are the generalized Bianchi identities and the law of
energy-momentum conservation.
The Lagrangian density is expanded to second-order, and treated as an
``Einstein plus fields'' theory.
{}From this, it is deduced that the energy is positive in the radiation
zone.

\end{abstract}


\narrowtext

\section{Introduction}

Recently, a consistent version of the nonsymmetric
gravitational theory (NGT) has been
proposed \cite{bib:Moffat_NGT_1,bib:Moffat_NGT_2}.
This theory is free of ghosts, tachyons and higher-order poles in the
propagator in the linear approximation \cite{bib:Moffat_NGT_1}.

In the following, we will present a detailed derivation of the field
equations and compatibility conditions for the NGT, starting from a
Lagrangian density.
Using the general covariance of this Lagrangian density, we will
deduce the conservation laws and tensor identities present in the theory.
These will be seen to be direct generalizations of their general
relativistic counterparts.

Finally, by expanding the Lagrangian density to second-order about an
arbitrary Einstein background, we will demonstrate that the
energy contributions of the NGT vanish for large $r$, leaving only
the contributions from general relativity (GR).
Since these are known to be positive-definite, we will conclude that
for large $r$, there are no negative energy modes in the NGT.

\section{Structure of the Nonsymmetric Gravitational Theory}

The NGT is a geometric theory of
gravity based on a nonsymmetric field structure:
$ g_{\mu\nu} = g_{(\mu\nu)} + g_{[\mu\nu]} $;
in the NGT, $ g_{[\mu\nu]} $ does not vanish.
The affine connection coefficients, $ \Gamma^\lambda_{\mu\nu} $,
are also nonsymmetric.
We define the inverse tensor $ g^{\mu\nu} $ by the relation
\[
g^{\mu\nu} g_{\mu\alpha} = g^{\nu\mu} g_{\alpha\mu} = \delta^\nu_\alpha .
\]

The Lagrangian density for the NGT can be written as the sum of four
contributions: $ {\cal L}_{\rm NGT} = {\cal L}_{\rm geom} +
{\cal L}_{\rm cosmo} + {\cal L}_{\rm skew} + {\cal L}_{W} $.
The geometric and cosmological
terms, $ {\cal L}_{\rm geom} $ and $ {\cal L}_{\rm cosmo} $, are defined
by analogy with their counterparts in GR:
$ {\cal L}_{\rm geom} = {\bf g}^{\mu\nu} R_{\mu\nu}(W) $
and
$ {\cal L}_{\rm cosmo} = - 2 \lambda \sqrt{-g} $ .
The remaining terms are defined by
\[
{\cal L}_{\rm skew} = - \frac{1}{4} \mu^2 {\bf g}^{\mu\nu} g_{[\nu\mu]} ,
\]
and
\[
{\cal L}_{W} = \frac{1}{2} \sigma {\bf g}^{(\mu\nu)} W_\mu W_\nu .
\]
$ \lambda $ and $ \mu^2 $ are two cosmological constants, while $ \sigma $ is a
coupling constant.
In the linearized theory, it is found that $ \sigma = -1/3 $.
The NGT Ricci curvature tensor $ R_{\mu\nu}(W) $ is given by
\begin{equation}
\label{eq:Ricci_in_W}
R_{\mu\nu}(W) = W^\beta_{\mu\nu,\beta}
- \frac{1}{2} ( W^\beta_{\mu\beta,\nu} + W^\beta_{\nu\beta,\mu} )
- W^\beta_{\alpha\nu} W^\alpha_{\mu\beta}
+ W^\beta_{\alpha\beta} W^\alpha_{\mu\nu} ,
\end{equation}
where the $ W^\lambda_{\mu\nu} $ are the
unconstrained nonsymmetric connection coefficients, defined in terms
of the affine connection coefficients through the relation:
\begin{equation}
\label{eq:definition_of_W}
W^\lambda_{\mu\nu} = \Gamma^\lambda_{\mu\nu}
- \frac{2}{3} \delta^\lambda_\mu W_\nu ,
\end{equation}
where $ W_\mu = W^\alpha_{[\mu\alpha]} $.
It follows from (\ref{eq:definition_of_W}) that
$ \Gamma_\mu = \Gamma^\lambda_{[\mu\lambda]} = 0 $.
The NGT Ricci scalar is given by $ R(W) = g^{\mu\nu}R_{\mu\nu}(W) $.

\section{Derivation of the Field Equations}

The action principle reads
\begin{equation}
\label{eq:action_principle}
\delta S = \delta \int ( {\cal L}_{\rm NGT} + {\cal L}_{\rm M} )
\, d^4 x = 0 ,
\end{equation}
where $ {\cal L}_{\rm M} $ is a matter coupling term, for which
\begin{equation}
\label{eq:matter_coupling}
\delta S_{\rm M} = \delta \int {\cal L}_{\rm M} \, d^4 x
= -8 \pi \int {\bf T}_{\mu\nu} \, \delta g^{\mu\nu} \, d^4 x .
\end{equation}
All variations are with respect to the $ g^{\mu\nu} $.

Note that
\[
\delta ( {\cal L}_{\rm geom} + {\cal L}_{\rm cosmo} ) =
\delta ( {\bf g}^{\mu\nu} R_{\mu\nu}(W) -2 \lambda \sqrt{-g} ) =
( {\bf G}_{\mu\nu}(W) + \lambda {\bf g}_{\mu\nu} ) \, \delta g^{\mu\nu} ,
\]
where
\[
G_{\mu\nu} = R_{\mu\nu} - \frac{1}{2} g_{\mu\nu} R
\]
as in GR.

Next,
\[
\delta {\cal L}_{\rm skew} =
- \frac{1}{4} \mu^2 \delta ( {\bf g}^{\mu\nu} g_{[\nu\mu]} )
= \frac{1}{4} \mu^2
\left( \frac{1}{2} {\bf g}_{\mu\nu} g^{[\alpha\beta]} g_{[\beta\alpha]}
+ {\bf g}_{[\mu\nu]}
+ {\bf g}^{[\alpha\beta]} g_{\mu\alpha} g_{\beta\nu}
\right) \delta g^{\mu\nu} .
\]
The parenthesized quantity is defined as $ {\bf C}_{\mu\nu} $, leaving
\[
\delta {\cal L}_{\rm skew} =
- \frac{1}{4} \mu^2 \delta ({\bf g}^{\mu\nu} g_{[\nu\mu]}) =
\frac{1}{4} \mu^2 {\bf C}_{\mu\nu} \, \delta g^{\mu\nu} .
\]

Finally,
\[
\delta {\cal L}_{W} =
\frac{1}{2} \sigma \delta ( {\bf g}^{\mu\nu} W_\mu W_\nu )
= \frac{1}{2} \sigma \sqrt{-g} \left( W_\mu W_\nu - \frac{1}{2} g_{\mu\nu}
g^{\alpha\beta} W_\alpha W_\beta \right) \delta g^{\mu\nu} .
\]
Defining $ P_{\mu\nu} = W_\mu W_\nu $ and $ P = g^{\mu\nu} P_{\mu\nu} $, we
have that
\[
\delta {\cal L}_{W} =
\frac{1}{2} \sigma \delta ( {\bf g}^{\mu\nu} W_\mu W_\nu )
= \frac{1}{2} \sigma
\left( {\bf P}_{\mu\nu} - \frac{1}{2} g_{\mu\nu} {\bf P} \right)
\delta g^{\mu\nu}
= \frac{1}{2} \sigma {\tilde {\bf P}_{\mu\nu}} \, \delta g^{\mu\nu} ,
\]
where
\[
{\tilde P_{\mu\nu}} = P_{\mu\nu} - \frac{1}{2}g_{\mu\nu}P .
\]

Assembling these results and using (\ref{eq:matter_coupling}), we find that
\[
\delta S = \int \sqrt{-g} \left(
G_{\mu\nu}(W) + \lambda g_{\mu\nu} + \frac{1}{4} \mu^2 C_{\mu\nu}
+ \frac{1}{2} \sigma {\tilde P_{\mu\nu}}
- 8 \pi T_{\mu\nu} \right) \delta g^{\mu\nu} \, d^4x = 0 .
\]
This must hold for arbitrary $ \delta g^{\mu\nu} $, yielding the NGT field
equations:
\begin{equation}
\label{eq:NGT_field_equations}
G_{\mu\nu}(W) + \lambda g_{\mu\nu} + \frac{1}{4} \mu^2 C_{\mu\nu}
+ \frac{1}{2} \sigma {\tilde P_{\mu\nu}}
= 8 \pi T_{\mu\nu} .
\end{equation}

\section{Compatibility Conditions}

Varying the action with respect to the $ W_\mu $ yields the compatibility
conditions for the NGT.
There are two contributions to consider:
\[
S_{\rm geom} = \int {\bf g}^{\mu\nu} R_{\mu\nu}(W) \, d^4 x
\]
and
\[
S_{W} = \frac{1}{2} \sigma \int {\bf g}^{\mu\nu} W_\mu W_\nu \, d^4 x .
\]
We consider these separately.

Consider first the variation of $ S_{W} $:
\[
\delta S_{W} = \sigma \int {\bf g}^{(\mu\nu)} W_\mu \,
\delta W_\nu \, d^4 x .
\]
Now,
\[
\delta W_\nu = \frac{1}{2} \left( \delta^\alpha_\eta \delta^\rho_\alpha
\, \delta W^\eta_{\nu\rho} - \delta^\alpha_\eta \delta^\rho_\alpha
\, \delta W^\eta_{\rho\nu} \right) .
\]
Therefore,
\[
\delta S_{W} = \frac{1}{2} \sigma \int \left(
{\bf g}^{(\mu\rho)} W_\mu  \delta^\sigma_\eta -
{\bf g}^{(\mu\sigma)} W_\mu  \delta^\rho_\eta \right)
\, \delta W^\eta_{\rho\sigma}
\, d^4 x .
\]

Consider now the variation of $ S_{\rm geom} $:
using (\ref{eq:Ricci_in_W}), we have
\begin{eqnarray*}
\delta S_{\rm geom}
= \int {\bf g}^{\mu\nu} \, \delta R_{\mu\nu}(W) \, d^4x
&=& \int {\bf g}^{\mu\nu}
\biggl[ \delta W^\beta_{\mu\nu,\beta}
- \frac{1}{2} \left( \delta W^\beta_{\mu\beta,\nu}
+ \delta W^\beta_{\nu\beta,\mu} \right) \\
& & \mbox{} - \delta W^\beta_{\alpha\nu} W^\alpha_{\mu\beta}
- W^\beta_{\alpha\nu} \delta W^\alpha_{\mu\beta}
+ \delta W^\beta_{\alpha\beta} W^\alpha_{\mu\nu}
+ W^\beta_{\alpha\beta} \delta W^\alpha_{\mu\nu} \biggr] \, d^4 x .
\end{eqnarray*}
Integrating the first three terms by parts and relabeling indices, we arrive at
\begin{eqnarray}
\delta S_{\rm geom}
&=& \int \biggl[ - {{\bf g}^{\rho\sigma}}_{,\eta}
+ \frac{1}{2} \left( {{\bf g}^{\rho\nu}}_{,\nu}
+ {{\bf g}^{\mu\rho}}_{,\mu} \right)
\delta^\sigma_\eta
- {\bf g}^{\mu\sigma} W^\rho_{\mu\eta}
- {\bf g}^{\rho\nu} W^\sigma_{\eta\nu} \nonumber \\
& & \label{eq:functional_derivative_of_L_geom}
\mbox{} + {\bf g}^{\mu\nu} W^\rho_{\mu\nu}\delta^\sigma_\eta
+ {\bf g}^{\rho\sigma} W^\beta_{\eta\beta} \biggr]
\, \delta W^\eta_{\rho\sigma} \, d^4 x ,
\end{eqnarray}
where we have assumed that the $ \delta W^\eta_{\rho\sigma} $ vanish on the
boundary of integration.

If we require that these variations vanish, we have that
\begin{eqnarray*}
0 = \delta S_{\rm geom} + \delta S_{W}
&=& \int \biggl[ - {{\bf g}^{\rho\sigma}}_{,\eta}
+ \frac{1}{2} \left( {{\bf g}^{\rho\nu}}_{,\nu}
+ {{\bf g}^{\mu\rho}}_{,\mu} \right)
\delta^\sigma_\eta
- {\bf g}^{\mu\sigma} W^\rho_{\mu\eta}
- {\bf g}^{\rho\nu} W^\sigma_{\eta\nu} \\
& & \mbox{} + {\bf g}^{\mu\nu} W^\rho_{\mu\nu} \delta^\sigma_\eta
+ {\bf g}^{\rho\sigma} W^\beta_{\eta\beta}
+ \frac{1}{2} \sigma \left(
{\bf g}^{(\mu\rho)} W_\mu \delta^\sigma_\eta -
{\bf g}^{(\mu\sigma)} W_\mu \delta^\rho_\eta \right) \biggr]
\, \delta W^\eta_{\rho\sigma} \, d^4x .
\end{eqnarray*}
Since this must hold for arbitrary $ \delta W^\eta_{\rho\sigma} $, we
arrive at
the compatibility conditions for the NGT:
\begin{equation}
\label{eq:full_compatibility}
{{\bf g}^{\rho\sigma}}_{,\eta}
- {{\bf g}^{(\rho\nu)}}_{,\nu} \delta^\sigma_\eta
+ {\bf g}^{\mu\sigma} W^\rho_{\mu\eta} + {\bf g}^{\rho\nu} W^\sigma_{\eta\nu}
- {\bf g}^{\mu\nu} W^\rho_{\mu\nu} \delta^\sigma_\eta
- {\bf g}^{\rho\sigma} W^\beta_{\eta\beta}
- \sigma {\bf g}^{(\mu\lambda)} W_\mu \delta^\rho_{[\lambda}
\delta^\sigma_{\eta{}]} = 0 .
\end{equation}
Contracting this on $ \rho $ and $ \eta $ gives
\begin{equation}
\label{eq:skew_compatibility}
{{\bf g}^{[\sigma\rho]}}_{,\rho} = \frac{3}{2} \sigma
{\bf g}^{(\rho\sigma)} W_\rho .
\end{equation}
Contracting (\ref{eq:full_compatibility}) on $ \sigma $ and $ \eta $,
and adding this to (\ref{eq:skew_compatibility}) gives
\[
{{\bf g}^{(\sigma\rho)}}_{,\rho} + {\bf g}^{\mu\rho} W^\sigma_{\mu\rho}
= \frac{2}{3} {\bf g}^{\sigma\nu} W^\rho_{[\rho\nu]}
= - \frac{2}{3} {\bf g}^{\sigma\nu} W_\nu .
\]
This may be used to rewrite (\ref{eq:full_compatibility}) as
\begin{equation}
\label{eq:full_compatibility_rewritten}
{{\bf g}^{\rho\sigma}}_{,\eta}
+ {\bf g}^{\mu\sigma} W^\rho_{\mu\eta} + {\bf g}^{\rho\nu} W^\sigma_{\eta\nu}
- {\bf g}^{\rho\sigma} W^\beta_{\eta\beta}
- \sigma {\bf g}^{(\mu\lambda)} W_\mu \delta^\rho_{[\lambda}
\delta^\sigma_{\eta{}]}
- \frac{2}{3} {\bf g}^{\rho\mu} W^\nu_{[\nu\mu]} \delta^\sigma_\eta = 0 .
\end{equation}

Inserting the expression for $ \Gamma^\lambda_{\mu\nu} $
obtained from (\ref{eq:definition_of_W}) into
(\ref{eq:full_compatibility_rewritten}) gives the compatibility
condition for the $ \Gamma^\lambda_{\mu\nu} $:
\begin{equation}
\label{eq:Gamma_compatibility}
g_{\lambda\xi,\eta} - g_{\rho\xi} \Gamma^\rho_{\lambda\eta}
- g_{\lambda\sigma} \Gamma^\sigma_{\eta\xi}
+ \frac{1}{2} \sigma g^{(\mu\rho)}
\left( g_{\rho\xi} g_{\lambda\eta} - g_{\eta\xi} g_{\lambda\rho}
- g_{\lambda\xi} g_{[\rho\eta]} \right) W_\mu = 0 .
\end{equation}

\section{Conservation Laws and Identities}

We now proceed to derive the conservation laws and tensor identities
present in the NGT
(see for example \cite{bib:Papapetrou}).

Every term in the NGT Lagrangian density $ {\cal L}_{\rm NGT} $
is a scalar density.
It follows that each term in the NGT action,
\[
S_{\rm NGT} = \int {\cal L}_{\rm NGT} \, d^4 x ,
\]
must be invariant under a general coordinate transformation.
In particular, we consider the infinitesimal coordinate transformation
generated by $ x^\mu \rightarrow x'^\mu = x^\mu + \epsilon\xi^\mu $, where
$ \epsilon \ll 1 $.
Since $ g_{\mu\nu} $ is a tensor and $ W_\mu $ is a vector, we have that
\begin{eqnarray*}
g'_{\mu\nu}(x)
&=& \frac{\partial x^\rho}{\partial x'^\mu}
\frac{\partial x^\sigma}{\partial x'^\nu}
g_{\rho\sigma}(x) + g'_{\mu\nu}(x) - g'_{\mu\nu}(x') \\
W'_\mu(x)
&=& \frac{\partial x^\rho}{\partial x'^\mu} W_\rho(x) + W'_\mu(x)
- W'_\mu(x') .
\end{eqnarray*}
It follows that
\begin{mathletters}
\label{eq:deltas_g_and_W}
\begin{eqnarray}
\label{eq:delta_g}
\delta g^{\mu\nu}
= - g^{\mu\beta} g^{\alpha\nu} ( g'_{\alpha\beta}(x) - g_{\alpha\beta}(x) )
&=& \epsilon \left( g^{\mu\rho} {\xi^\nu}_{,\rho}
+ g^{\rho\nu} {\xi^\mu}_{,\rho}
- {g^{\mu\nu}}_{,\lambda} \xi^\lambda \right) \\
\label{eq:delta_W}
\delta W_\mu = W'_\mu(x) - W_\mu(x)
&=& - \epsilon \left( W_\lambda {\xi^\lambda}_{,\mu}
+ W_{\mu,\lambda} \xi^\lambda \right) ,
\end{eqnarray}
\end{mathletters}to first order in $ \epsilon $.
Here, we have used the fact that
$ g^{\rho\nu} g^{\mu\sigma} g_{\rho\sigma,\lambda}
= - {g^{\mu\nu}}_{,\lambda} $.

Consider first the term
$ {\cal L}_{\rm geom} = {\bf g}^{\mu\nu} R_{\mu\nu}(W) $ appearing in the
NGT Lagrangian density.
Since this is a scalar density, we must necessarily have
\begin{equation}
\label{eq:first_variation}
\delta S_{\rm geom} = 0
= \int \left( \frac{\delta {\cal L}_{\rm geom}}{\delta g^{\mu\nu}}
\delta g^{\mu\nu}
+ \frac{\delta {\cal L}_{\rm geom}}{\delta W^\eta_{\rho\sigma}}
\delta W^\eta_{\rho\sigma} \right) \, d^4 x ,
\end{equation}
where $ \delta / \delta g^{\mu\nu} $ denotes functional differentiation.
The second term was evaluated in (\ref{eq:functional_derivative_of_L_geom});
using the compatibility condition, this can be reduced to
\[
\frac{\delta {\cal L}_{\rm geom}}{\delta W^\eta_{\rho\sigma}}
\delta W^\eta_{\rho\sigma}
= \sigma {\bf g}^{(\mu\lambda)} W_\mu \delta^\sigma_{[\lambda}
\delta^\rho_{\eta]} \, \delta W^\eta_{\rho\sigma}
= - \sigma {\bf g}^{(\mu\lambda)} W_\mu \, \delta W_\lambda ,
\]
where we have used the fact that
$ \delta W^\eta_{[\eta\lambda]} = - \delta W_\lambda $.
This can be further simplified through the use of
(\ref{eq:skew_compatibility}), yielding
\[
\frac{\delta {\cal L}_{\rm geom}}{\delta W^\eta_{\rho\sigma}}
\delta W^\eta_{\rho\sigma}
= - \frac{2}{3} {{\bf g}^{[\lambda\mu]}}_{,\mu} \, \delta W_\lambda .
\]

As was found in the derivation of the field equations, the first term
in (\ref{eq:first_variation}) is given by
\[
\frac{\delta {\cal L}_{\rm geom}}{\delta g^{\mu\nu}} \delta g^{\mu\nu}
= {\bf G}_{\mu\nu}(W) \, \delta g^{\mu\nu} .
\]
Combining these results and using (\ref{eq:deltas_g_and_W}), we have that
\begin{eqnarray}
\delta S_{\rm geom}
&=& \epsilon \int {\bf G}_{\mu\nu}(W)
\left( g^{\mu\rho} {\xi^\nu}_{,\rho} + g^{\rho\nu} {\xi^\mu}_{,\rho}
- {g^{\mu\nu}}_{,\rho} \xi^\rho \right) \, d^4x
\mbox{} + \frac{2}{3} \epsilon
\int {{\bf g}^{[\lambda\mu]}}_{,\mu} \left( W_\rho {\xi^\rho}_{,\lambda}
+ W_{\lambda,\rho} \xi^\rho \right) \, d^4x \nonumber \\
&=& - \epsilon \int \left( \left[ g^{\mu\rho} {\bf G}_{\mu\sigma}(W)
+ g^{\rho\nu} {\bf G}_{\sigma\nu}(W) \right]_{,\rho}
+ {g^{\mu\nu}}_{,\sigma} {\bf G}_{\mu\nu}(W)
- \frac{4}{3} {{\bf g}^{[\rho\mu]}}_{,\mu} W_{[\rho,\sigma]} \right)
\xi^\sigma \, d^4x \nonumber \\
& & \label{eq:general_variation}
\mbox{} + \epsilon \int \left[ \left(
g^{\mu\rho} {\bf G}_{\mu\sigma}(W) + g^{\rho\nu} {\bf G}_{\sigma\nu}(W)
+ \frac{2}{3} {{\bf g}^{[\rho\mu]}}_{,\mu} W_\sigma \right) \xi^\sigma
\right]_{,\rho} d^4x .
\end{eqnarray}

Suppose first that the $ \xi^\sigma $ vanish on the boundary
of integration, but are otherwise arbitrary.
The second term of (\ref{eq:general_variation}) must vanish, since
it is strictly a surface term.
The first term yields
\begin{equation}
\label{eq:Bianchi_in_W}
\left[ g^{\mu\rho}{\bf G}_{\mu\sigma}(W)
+ g^{\rho\mu}{\bf G}_{\sigma\mu}(W)\right]_{,\rho}
+ {g^{\mu\rho}}_{,\sigma}{\bf G}_{\mu\rho}(W)
- \frac{4}{3} {{\bf g}^{[\rho\mu]}}_{,\mu} W_{[\rho,\sigma]} = 0 .
\end{equation}
These are known as the generalized Bianchi identities of the NGT.
They can be written in terms of the $ G_{\mu\nu}(\Gamma) $ by first noting
that
\[
G_{\mu\nu}(W) = G_{\mu\nu}(\Gamma)
+ \frac{2}{3} \left( W_{[\mu,\nu]}
- \frac{1}{2} g_{\mu\nu} g^{[\alpha\beta]} W_{[\alpha,\beta]} \right) .
\]
By direct substitution into (\ref{eq:Bianchi_in_W}), it is seen that
\[
\left[ g^{\mu\rho} {\bf G}_{\mu\sigma}(\Gamma)
+ g^{\rho\mu} {\bf G}_{\sigma\mu}(\Gamma) \right]_{,\rho}
+ {g^{\mu\nu}}_{,\sigma} {\bf G}_{\mu\nu}(\Gamma)
= \frac{2}{3} {\bf g}^{[\mu\nu]} W_{[\mu,\nu],\sigma}
- \frac{4}{3} {\bf g}^{[\mu\nu]} W_{[\mu,\sigma],\nu} .
\]
However,
\[
{\bf g}^{[\mu\nu]} W_{[\mu,\sigma],\nu}
= \frac{1}{2} {\bf g}^{[\mu\nu]} W_{[\mu,\nu],\sigma} .
\]
Therefore, in terms of the $ G_{\mu\nu}(\Gamma) $, the Bianchi identities
are written
\begin{equation}
\label{eq:Bianchi}
\left[ g^{\mu\rho} {\bf G}_{\mu\sigma}(\Gamma)
+ g^{\rho\mu} {\bf G}_{\sigma\mu}(\Gamma) \right]_{,\rho}
+ {g^{\mu\nu}}_{,\sigma} {\bf G}_{\mu\nu}(\Gamma) = 0 .
\end{equation}

Inserting (\ref{eq:Bianchi_in_W}) into (\ref{eq:general_variation}) leaves
\[
\epsilon \int \left[ \left( g^{\mu\rho} {\bf G}_{\mu\sigma}(W)
+ g^{\rho\mu} {\bf G}_{\sigma\mu}(W)
+ \frac{2}{3} {{\bf g}^{[\rho\mu]}}_{,\mu} W_\sigma \right)
\xi^\sigma \right]_{,\rho} d^4x = 0 .
\]
Suppose $\xi^\rho$ is taken as an arbitrary constant vector;
we then have
\[
\epsilon \int \left[ g^{\mu\rho} {\bf G}_{\mu\sigma}(W)
+ g^{\rho\mu} {\bf G}_{\sigma\mu}(W)
+ \frac{2}{3} {{\bf g}^{[\rho\mu]}}_{,\mu} W_\sigma
\right]_{,\rho} \xi^\sigma d^4x = 0 .
\]
We can use (\ref{eq:Bianchi_in_W}) to simplify this, leaving
\[
\epsilon \int \left( {g^{\mu\rho}}_{,\sigma} {\bf G}_{\mu\rho}(W)
- \frac{2}{3} {{\bf g}^{[\rho\mu]}}_{,\mu} W_{\rho,\sigma} \right)
\xi^\sigma d^4x = 0 .
\]
Therefore,
\begin{equation}
\label{eq:other_identity}
{g^{\mu\rho}}_{,\sigma} {\bf G}_{\mu\rho}(W)
= \frac{2}{3} {{\bf g}^{[\rho\mu]}}_{,\mu} W_{\rho,\sigma} .
\end{equation}
The identities (\ref{eq:Bianchi_in_W}) and
(\ref{eq:other_identity}) are all the relations that may be derived
from the general covariance of $ {\cal L}_{\rm geom} $.

Inserting the field equations into (\ref{eq:Bianchi}) gives
\begin{equation}
\label{eq:matter_response}
g_{\mu\lambda}{{\bf T}^{*\mu\rho}}_{,\rho}
+ g_{\lambda\mu}{{\bf T}^{*\rho\mu}}_{,\rho}
+ (g_{\mu\lambda,\nu}
+ g_{\lambda\nu,\mu}
- g_{\mu\nu,\lambda}){\bf T}^{*\mu\nu}
- \frac{1}{4\pi}\sigma{\bf g}^{(\mu\nu)}W_\nu W_{[\mu,\lambda]} = 0 ,
\end{equation}
where we have introduced
\[
T^*_{\mu\nu} = T_{\mu\nu} - \frac{1}{32\pi} \mu^2 C_{\mu\nu}
- \frac{1}{16\pi} \sigma {\tilde P_{\mu\nu}}
\]
for brevity.

Consider now the two terms
\[
{\cal L}_{\rm skew} = -\frac{1}{4}\mu^2 {\bf g}^{\mu\nu} g_{[\nu\mu]}
\]
and
\[
{\cal L}_{W} = \frac{1}{2}\sigma {\bf g}^{(\mu\nu)} W_\mu W_\nu
\]
that appear in the NGT Lagrangian density.
Both these terms are scalar densities, and hence their corresponding
contributions to the action must also be invariant under the transformation,
$x^\mu \rightarrow x'^\mu = x^\mu+\epsilon\xi^\mu$, considered above.
Proceeding in the same manner as for the curvature term,
these two contributions lead to the identities
\begin{mathletters}
\label{eq:tensor_identities}
\begin{equation}
g_{\mu\lambda}{{\bf C}^{\mu\rho}}_{,\rho}
+ g_{\lambda\mu}{{\bf C}^{\rho\mu}}_{,\rho}
+ (g_{\mu\lambda,\nu}
+ g_{\lambda\nu,\mu}
- g_{\mu\nu,\lambda}){\bf C}^{\mu\nu} = 0
\end{equation}
and
\begin{equation}
g_{\mu\lambda}{\tilde {\bf P}^{\mu\rho}}{}_{,\rho}
+ g_{\lambda\mu}{\tilde {\bf P}^{\rho\mu}}{}_{,\rho}
+ (g_{\mu\lambda,\nu}
+ g_{\lambda\nu,\mu}
- g_{\mu\nu,\lambda})\tilde {\bf P}^{\mu\nu}
+ 4{\bf g}^{(\mu\nu)} W_\mu W_{[\nu,\lambda]} = 0 .
\end{equation}
\end{mathletters}However, the identities appearing
in (\ref{eq:tensor_identities})
are two of the terms that appear in
(\ref{eq:matter_response}).
Cancelling these terms, we arrive at
\begin{equation}
\label{eq:energy-momentum_conserve}
g_{\mu\lambda}{{\bf T}^{\mu\rho}}_{,\rho}
+ g_{\lambda\mu}{{\bf T}^{\rho\mu}}_{,\rho}
+ (g_{\mu\lambda,\nu} + g_{\lambda\nu,\mu} - g_{\mu\nu,\lambda})
{\bf T}^{\mu\nu} = 0 .
\end{equation}
This is known as the generalized law of energy-momentum conservation in NGT.
This is a direct generalization of the identity
$ \nabla_\nu T^{\mu\nu} = 0 $ of GR, where $\nabla_\nu$ denotes covariant
differentiation with respect to the GR connection.
As a matter of fact, if
$ g_{\mu\nu} $ is taken as symmetric, this reduces
to $\nabla_\nu T^{(\mu\nu)} = 0 $.

\section{Positivity of the Asymptotic Value of the Energy}

We now consider a second-order expansion of the NGT Lagrangian density about
an arbitrary Einstein background.
To this end, we write
\begin{eqnarray*}
g_{\mu\nu} &=& {}^{({\rm E})} \! g_{\mu\nu} + {}^{(1)} \! g_{\mu\nu}
+ {}^{(2)} \! g_{\mu\nu} + \ldots \\
\Gamma^\lambda_{\mu\nu} &=& {}^{({\rm E})} \Gamma^\lambda_{\mu\nu}
+ {}^{(1)} \Gamma^\lambda_{\mu\nu}
+ {}^{(2)} \Gamma^\lambda_{\mu\nu} + \ldots \\
W_\mu &=& {}^{(1)} W_\mu + {}^{(2)} W_\mu + \ldots
\end{eqnarray*}
Here and throughout, we use the convention that a quantity preceded
by a $ {}^{({\rm E})} $ is to be evaluated in the Einstein background.
Since $ W_\mu $ has no equivalent in GR, there is no
$ {}^{({\rm E})} W_\mu $ term.
The inverse metric $ g^{\mu\nu} $ is given by
\[
g^{\mu\nu} = {}^{({\rm E})} \! g^{\mu\nu} + {}^{(1)} \! g^{\mu\nu}
+ {}^{(2)} \! g^{\mu\nu} + \ldots ,
\]
where $ {}^{({\rm E})} \! g^{\mu\nu} $ is the usual inverse metric from
GR, and
\begin{eqnarray*}
{}^{(1)} \! g^{\beta\nu} &=& - {}^{({\rm E})} \! g^{\beta\alpha}
{}^{({\rm E})} \! g^{\mu\nu} {}^{(1)} \! g_{\mu\alpha} \\
{}^{(2)} \! g^{\beta\nu} &=& - {}^{({\rm E})} \! g^{\beta\alpha}
{}^{({\rm E})} \! g^{\mu\nu} {}^{(2)} \! g_{\mu\alpha}
- {}^{({\rm E})} \! g^{\beta\alpha} {}^{(1)} \! g^{\mu\nu}
{}^{(1)} \! g_{\mu\alpha} .
\end{eqnarray*}
To second-order, the determinant of the metric is
$ g = {}^{({\rm E})} \! g
+ {}^{({\rm E})}\!g {}^{(1)} \! g
+ {}^{({\rm E})}\!g {}^{(2)} \! g + \ldots $, where
\begin{eqnarray*}
{}^{(1)} \! g &=& {}^{({\rm E})} \! g_{\alpha\mu}
{}^{(1)} \! g^{\alpha\mu} \\
{}^{(2)} \! g &=& \frac{3}{8}
\left( {}^{({\rm E})} \! g_{\alpha\mu}
{}^{(1)} \! g^{\alpha\mu} \right)^2
+ {}^{({\rm E})} \! g_{\alpha\mu} {}^{(2)} \! g^{\alpha\mu} .
\end{eqnarray*}
It therefore follows that, to the same order of approximation,
\[
\sqrt{-g}
= \sqrt{-{}^{({\rm E})} \! g} \left[
1 + \frac{1}{2} {}^{({\rm E})} \! g_{\mu\nu} {}^{(1)} \! g^{\mu\nu}
- \frac{1}{16} \left( {}^{({\rm E})} \! g_{\mu\nu}
{}^{(1)} \! g^{\mu\nu} \right)^2
+ \frac{1}{2} {}^{({\rm E})} \! g_{\mu\nu} {}^{(2)} \! g^{\mu\nu} \right] .
\]

If we expand the compatibility condition (\ref{eq:Gamma_compatibility})
to lowest-order, we obtain
\[
{}^{({\rm E})} \! g_{\lambda\xi,\eta} -
{}^{({\rm E})} \! g_{\rho\xi} {}^{({\rm E})} \Gamma^\rho_{\lambda\eta} -
{}^{({\rm E})} \! g_{\lambda\rho} {}^{({\rm E})} \Gamma^\rho_{\eta\xi} = 0 .
\]
This is recognized as the compatibility condition familiar from
GR.
Its solution is well-known:
\[
{}^{({\rm E})} \Gamma^\sigma_{\lambda\eta}
= \frac{1}{2} {}^{({\rm E})} \! g^{\sigma\xi} \left(
{}^{({\rm E})} \! g_{\lambda\xi,\eta} + {}^{({\rm E})} \! g_{\xi\eta,\lambda}
- {}^{({\rm E})} \! g_{\eta\lambda,\xi} \right) .
\]
The first- and second-order corrections are obtained by a process of
iteration.
The results are
\begin{mathletters}
\label{eq:affine_corrections}
\begin{eqnarray}
{}^{(1)} \Gamma^\sigma_{\lambda\eta}
&=& \frac{1}{2} {}^{({\rm E})} \! g^{\sigma\xi} \left(
\nabla_\eta {}^{(1)} \! g_{\lambda\xi}
+ \nabla_\lambda {}^{(1)} \! g_{\xi\eta}
- \nabla_\xi {}^{(1)} \! g_{\eta\lambda} \right)
+ \sigma \delta^\sigma_{[\lambda} \delta^\mu_{\eta]}
{}^{(1)} W_\mu \\
{}^{(2)} \Gamma^\sigma_{\lambda\eta}
&=& \frac{1}{2} {}^{({\rm E})} \! g^{\sigma\xi} \left(
\nabla_\eta {}^{(2)} \! g_{\lambda\xi} + \nabla_\lambda
{}^{(2)} \! g_{\xi\eta}
- \nabla_\xi {}^{(2)} \! g_{\eta\lambda} \right)
+ \sigma\delta^\sigma_{[\lambda}\delta^\mu_{\eta]} {}^{(2)} W_\mu
\nonumber \\
& & \mbox{}
- {}^{({\rm E})} \! g^{\sigma\xi} {}^{(1)} \! g_{\rho\xi}
{}^{(1)} \Gamma^\rho_{\lambda\eta}
+ \frac{1}{2} \sigma \biggl[
{}^{({\rm E})} \! g^{\sigma\xi} \left(
{}^{(1)} \! g_{\xi\lambda} \delta^\mu_\eta
- {}^{(1)} \! g_{\eta\xi} \delta^\mu_\lambda \right)
{}^{(1)} W_\mu \nonumber \\
& & \mbox{}
- \frac{1}{2} {}^{({\rm E})} \! g^{\mu\rho}
\left( \delta^\sigma_\lambda {}^{(1)} \! g_{[\rho\eta]}
+ \delta^\sigma_\eta {}^{(1)} \! g_{[\rho\lambda]}
- {}^{({\rm E})} \! g^{\sigma\xi} {}^{({\rm E})} \! g_{\eta\lambda}
{}^{(1)} \! g_{[\rho\xi]} \right) {}^{(1)} W_\mu \biggr] .
\end{eqnarray}
\end{mathletters}Here, $ \nabla_\mu $
denotes covariant differentiation with respect to
the background metric $ {}^{({\rm E})} \! g_{\mu\nu} $.

The NGT Ricci curvature tensor $ R_{\mu\nu}(\Gamma) $ may be expanded in
a similar fashion.
Writing
\[
R_{\mu\nu}(\Gamma) =
{}^{({\rm E})} \! R_{\mu\nu}(\Gamma) +
{}^{(1)} \! R_{\mu\nu}(\Gamma) +
{}^{(2)} \! R_{\mu\nu}(\Gamma) + \ldots ,
\]
it is found that
\begin{mathletters}
\label{eq:curvature_expansions}
\begin{eqnarray}
{}^{({\rm E})} \! R_{\mu\nu}(\Gamma)
&=& {}^{({\rm E})} \Gamma^\beta_{\mu\nu,\beta}
- {}^{({\rm E})} \Gamma^\beta_{\mu\beta,\nu}
- {}^{({\rm E})} \Gamma^\beta_{\alpha\nu}
{}^{({\rm E})} \Gamma^\alpha_{\mu\beta}
+ {}^{({\rm E})} \Gamma^\beta_{\alpha\beta}
{}^{({\rm E})} \Gamma^\alpha_{\mu\nu} \nonumber \\
{}^{(1)} \! R_{\mu\nu}(\Gamma)
&=& \nabla_\beta {}^{(1)} \Gamma^\beta_{\mu\nu}
- \delta^\sigma_{(\nu} \delta^\rho_{\mu)}
\nabla_\sigma {}^{(1)} \Gamma^\beta_{(\rho\beta)} \\
{}^{(2)} \! R_{\mu\nu}(\Gamma)
&=& \nabla_\beta {}^{(2)} \Gamma^\beta_{\mu\nu}
- \delta^\sigma_{(\nu} \delta^\rho_{\mu)}
\nabla_\sigma {}^{(2)} \Gamma^\beta_{(\rho\beta)}
- {}^{(1)} \Gamma^\beta_{\alpha\nu} {}^{(1)} \Gamma^\alpha_{\mu\beta}
+ {}^{(1)} \Gamma^\beta_{(\alpha\beta)} {}^{(1)} \Gamma^\alpha_{\mu\nu} .
\end{eqnarray}
\end{mathletters}Note that $ {}^{({\rm E})} \! R_{\mu\nu}(\Gamma) $
is the usual Ricci
curvature tensor of GR, as expected.

In a second-order expansion, the NGT Lagrangian density is found to be
\[
{\cal L}_{\rm NGT} = {}^{({\rm E})} \! {\cal L} + {}^{(1)} \! {\cal L} +
{}^{(2)} \! {\cal L} + \ldots ,
\]
where
\[
{}^{({\rm E})} \! {\cal L}
= \sqrt{-{}^{({\rm E})} \! g} {}^{({\rm E})} \! g^{\mu\nu}
{}^{({\rm E})} \! R_{\mu\nu}(\Gamma)
- 2\lambda \sqrt{-{}^{({\rm E})} \! g}
\]
is the usual Lagrangian density of GR, and
\begin{mathletters}
\label{eq:Lagrangian_corrections}
\begin{eqnarray}
{}^{(1)} \! {\cal L}
&=& \sqrt{-{}^{({\rm E})} \! g} \left[ \frac{1}{2}
{}^{(1)} \! g
{}^{({\rm E})} \! g^{\mu\nu}
{}^{({\rm E})} \! R_{\mu\nu}(\Gamma)
+ {}^{(1)} \! g^{\mu\nu} {}^{({\rm E})} \! R_{\mu\nu}(\Gamma)
+ {}^{({\rm E})} \! g^{\mu\nu} {}^{(1)} \! R_{\mu\nu}(\Gamma)
- \lambda
{}^{(1)} \! g \right] \\
{}^{(2)} \! {\cal L}
&=& \sqrt{-{}^{({\rm E})} \! g} \Biggl\{ \Biggl[
\frac{1}{2}{}^{(2)} \! g
- \frac{1}{4} \left( {}^{(1)} \! g \right)^2
\Biggr] {}^{({\rm E})} \! g^{\mu\nu} {}^{({\rm E})} \! R_{\mu\nu}(\Gamma)
+ \frac{1}{2} {}^{(1)} \! g
{}^{(1)} \! g^{\mu\nu}
{}^{({\rm E})} \! R_{\mu\nu}(\Gamma) \nonumber \\
& & \mbox{} \vphantom{\Biggl(\Biggr)}
+ \frac{1}{2} {}^{(1)} \! g
{}^{({\rm E})} \! g^{\mu\nu}
{}^{(1)} \! R_{\mu\nu}(\Gamma)
+ {}^{(1)} \! g^{\mu\nu} {}^{(1)} \! R_{\mu\nu}(\Gamma)
+ {}^{(2)} \! g^{\mu\nu} {}^{({\rm E})} \! R_{\mu\nu}(\Gamma)
+ {}^{({\rm E})} \! g^{\mu\nu} {}^{(2)} \! R_{\mu\nu}(\Gamma)  \nonumber \\
& & \mbox{} \vphantom{\Biggl(\Biggr)}
- \lambda
\Biggl[ {}^{(2)} \! g
- \frac{1}{2} \left( {}^{(1)} \! g \right)^2
\Biggr]
- \frac{1}{4} \mu^2
{}^{(1)} g^{[\mu\nu]} {}^{(1)} \! g_{[\nu\mu]}
+ \frac{1}{2} \sigma
{}^{({\rm E})} \! g^{\mu\nu}
{}^{(1)} W_\mu {}^{(1)} W_\nu \nonumber \\
& & \mbox{} \vphantom{\Biggl(\Biggr)}
+ \frac{2}{3} {}^{(1)} \! g^{[\mu\nu]}
{}^{(1)} W_{[\mu,\nu]} \Biggr\} .
\end{eqnarray}
\end{mathletters}

We are now in a position to treat this problem as a special case of
GR.
Let us consider
\[
{\cal L}_{\rm NGT} = {}^{({\rm E})} \! {\cal L} + {\cal L}_{\rm field} ,
\]
where $ {\cal L}_{\rm field} = {}^{(1)} \! {\cal L} + {}^{(2)} \! {\cal L} $.
We can then define a stress-energy tensor (see \cite{bib:MTW})
by $ t^{\alpha\beta} =
{}^{(1)} t^{\alpha\beta} + {}^{(2)} t^{\alpha\beta} $, where
\[
{}^{(i)} {\bf t}^{\alpha\beta} \equiv
2 \frac{\delta \, {}^{(i)} \! {\cal L}}
{\delta \, {}^{({\rm E})} \! g_{\alpha\beta}} =
{}^{({\rm E})}\! g^{\alpha\beta}
{}^{(i)}\!{\cal L}
- 2\sqrt{-{}^{({\rm E})}\!g}
\frac{\delta\,{}^{(i)}\! L}{\delta\,{}^{({\rm E})}\! g_{\alpha\beta}} .
\]
The second term in ${}^{(i)}{\bf t}^{\alpha\beta}$ is given by
(\ref{eq:Lagrangian_corrections}).
The functional derivatives are found to be
\begin{eqnarray*}
\frac{\delta\,{}^{(1)}\!L}{\delta\,{}^{({\rm E})}\!g_{\alpha\beta}}
&=& \frac{1}{2}\left( - {}^{(1)}\!g^{\alpha\beta}{}^{({\rm E})}\!g^{\mu\nu}
{}^{({\rm E})}\!R_{\mu\nu}(\Gamma)
- {}^{(1)}\!g{}^{({\rm E})}\!g^{\mu\beta}{}^{({\rm E})}\!g^{\alpha\nu}
{}^{({\rm E})}\!R_{\mu\nu}(\Gamma)
+ {}^{(1)}\!g{}^{({\rm E})}\!g^{\mu\nu}
\frac{\delta\,{}^{({\rm E})}\!R_{\mu\nu}(\Gamma)}
{\delta\,{}^{({\rm E})}\! g_{\alpha\beta}}
\right) \\
& & \mbox{}
- \left( {}^{({\rm E})}\!g^{\mu\beta}{}^{(1)}\!g^{\alpha\nu}
+ {}^{({\rm E})}\!g^{\alpha\nu} {}^{(1)}\!g^{\mu\beta} \right)
{}^{({\rm E})}\! R_{\mu\nu}(\Gamma)
+ {}^{(1)}\!g^{\mu\nu}
\frac{\delta\,{}^{({\rm E})}\! R_{\mu\nu}(\Gamma)}
{\delta\,{}^{({\rm E})}\!g_{\alpha\beta}}
- {}^{({\rm E})}\!g^{\mu\beta}{}^{({\rm E})}\!g^{\alpha\nu}
{}^{(1)}\!R_{\mu\nu}(\Gamma) \\
& & \mbox{}
+ {}^{({\rm E})}\!g^{\mu\nu}
\frac{\delta\,{}^{(1)}\! R_{\mu\nu}(\Gamma)}
{\delta\,{}^{({\rm E})}\!g_{\alpha\beta}}
+ \lambda{}^{(1)}\!g^{\alpha\beta} \\
\frac{\delta\,{}^{(2)}\!L}{\delta\,{}^{({\rm E})}\!g_{\alpha\beta}}
&=& \frac{1}{2}\Biggl\{
\frac{\delta\,{}^{(2)}\!g}{\delta\,{}^{({\rm E})}\!g_{\alpha\beta}}
{}^{({\rm E})}\!g^{\mu\nu}{}^{({\rm E})}\!R_{\mu\nu}(\Gamma)
- {}^{(2)}\!g{}^{({\rm E})}\!g^{\mu\beta}{}^{({\rm E})}\!g^{\alpha\nu}
{}^{({\rm E})}\!R_{\mu\nu}(\Gamma)
+ {}^{(2)}\!g{}^{({\rm E})}\!g^{\mu\nu}
\frac{\delta\,{}^{({\rm E})}\!R_{\mu\nu}(\Gamma)}
{\delta\,{}^{({\rm E})}\!g_{\alpha\beta}} \\
& & \mbox{}
+ {}^{(1)}\!g{}^{(1)}\!g^{\alpha\beta}{}^{({\rm E})}\!g^{\mu\nu}
{}^{({\rm E})}\!R_{\mu\nu}(\Gamma)
+ \frac{1}{2}\left({}^{(1)}\!g\right)^2{}^{({\rm E})}\!g^{\mu\beta}
{}^{({\rm E})}\!g^{\alpha\nu}{}^{(2)}\!R_{\mu\nu}(\Gamma)
\phantom{
\frac{\delta\,{}^{({\rm E})}\!R_{\mu\nu}(\Gamma)}
{\delta\,{}^{({\rm E})}\!g_{\alpha\beta}}} \\
& & \mbox{}
- \frac{1}{2}\left({}^{(1)}\!g\right)^2{}^{({\rm E})}\!g^{\mu\nu}
\frac{\delta\,{}^{({\rm E})}\!R_{\mu\nu}(\Gamma)}
{\delta\,{}^{({\rm E})}\!g_{\alpha\beta}}
- {}^{(1)}\!g^{\alpha\beta}{}^{(1)}\!g^{\mu\nu}
{}^{({\rm E})}\!R_{\mu\nu}(\Gamma) \\
& & \mbox{}
- \left({}^{({\rm E})}\!g^{\mu\beta}{}^{(1)}\!g^{\alpha\nu}
+ {}^{({\rm E})}\!g^{\alpha\nu}{}^{(1)}\!g^{\mu\beta}\right)
{}^{(1)}\!g{}^{({\rm E})}\!R_{\mu\nu}(\Gamma)
+ {}^{(1)}\!g{}^{(1)}\!g^{\mu\nu}
\frac{\delta\,{}^{({\rm E})}\!R_{\mu\nu}(\Gamma)}
{\delta\,{}^{({\rm E})}\!g_{\alpha\beta}} \\
& & \mbox{}
- {}^{(1)}\!g^{\alpha\beta}{}^{({\rm E})}\!g^{\mu\nu}
{}^{(1)}\!R_{\mu\nu}(\Gamma)
- {}^{(1)}\!g{}^{({\rm E})}\!g^{\mu\beta}{}^{({\rm E})}\!g^{\alpha\nu}
{}^{(1)}\!R_{\mu\nu}(\Gamma)
+ {}^{(1)}\!g{}^{({\rm E})}\!g^{\mu\nu}
\frac{\delta\,{}^{(1)}\!R_{\mu\nu}(\Gamma)}
{\delta\,{}^{({\rm E})}\!g_{\alpha\beta}}
\Biggr\} \\
& & \mbox{}
- \left({}^{({\rm E})}\!g^{\mu\beta}{}^{(1)}\!g^{\alpha\nu}
+ {}^{({\rm E})}\!g^{\alpha\nu}{}^{(1)}\!g^{\mu\beta}\right)
{}^{(1)}\!R_{\mu\nu}(\Gamma)
+ {}^{(1)}\!g^{\mu\nu}
\frac{\delta\,{}^{(1)}\!R_{\mu\nu}(\Gamma)}
{\delta\,{}^{({\rm E})}\!g_{\alpha\beta}}
+ {}^{(2)}\!g^{\mu\nu}
\frac{\delta\,{}^{({\rm E})}\!R_{\mu\nu}(\Gamma)}
{\delta\,{}^{({\rm E})}\!g_{\alpha\beta}} \\
& & \mbox{}
- \left({}^{({\rm E})}\!g^{\mu\beta}{}^{(2)}\!g^{\alpha\nu}
+ {}^{({\rm E})}\!g^{\alpha\nu} {}^{(2)}\!g^{\mu\beta}
+ {}^{(1)}\!g^{\mu\beta} {}^{(1)}\!g^{\alpha\nu}\right)
{}^{({\rm E})}\!R_{\mu\nu}(\Gamma)
- {}^{({\rm E})}\!g^{\mu\beta}{}^{({\rm E})}\!g^{\nu\alpha}
{}^{(2)}\!R_{\mu\nu}(\Gamma)
\vphantom{
\frac{\delta\,{}^{({\rm E})}\!R_{\mu\nu}(\Gamma)}
{\delta\,{}^{({\rm E})}\!g_{\alpha\beta}}} \\
& & \mbox{}
+ {}^{({\rm E})}\!g^{\mu\nu}
\frac{\delta\,{}^{(2)}\!R_{\mu\nu}(\Gamma)}
{\delta\,{}^{({\rm E})}\!g_{\alpha\beta}}
- \lambda\left(
\frac{\delta\,{}^{(2)}\!g}{\delta\,{}^{({\rm E})}\!g_{\alpha\beta}}
+ {}^{(1)}\!g{}^{(1)}\!g^{\alpha\beta}\right)
- \frac{1}{2}\sigma{}^{({\rm E})}\!g^{\mu\beta}
{}^{({\rm E})}\!g^{\alpha\nu}
{}^{(1)}W_\mu {}^{(1)}W_\nu \\
& & \mbox{}
+ \left({}^{({\rm E})}\!g^{\mu\beta}{}^{(1)}\!g^{[\alpha\nu]}
+ {}^{({\rm E})}\!g^{\alpha\nu} {}^{(1)}\!g^{[\mu\beta]}\right)
\left(\frac{1}{4}\mu^2
{}^{(1)}\!g_{[\nu\mu]}
- \frac{2}{3}
{}^{(1)}W_{[\mu,\nu]}\right)
\end{eqnarray*}
The various terms in these expressions are given in Appendix
\ref{sec:appendix_expressions}.

Consider now the flux of stress-energy at infinity as given by the previous
two tensors.
For large $r$, it can be shown that ${}^{(i)} \! g_{\alpha\beta}$
(where $i=1,2$) is damped
out; it follows that, even in the worst possible case,
${}^{(i)} \Gamma^\lambda_{\mu\nu}$ and
${}^{(i)} \! R_{\mu\nu}(\Gamma)$ will decay at least as fast as
${}^{(i)} \! g_{\alpha\beta}$.
It has been shown (see \cite{bib:Neil}) that for large $r$,
\[
{}^{(i)} \! g_{[23]} \sim C
\sin\theta
\frac{e^{-\mu r}(1+\mu r)}{(\mu r)^{\mu M}} ,
\]
where $C$ is a constant and $M$ is the mass of the gravitating body.
In \cite{bib:Moffat_NGT_2}, it was shown that in the expansion to
linear order about an arbitrary Einstein background (with $\sigma=-1/3$):
\[
{}^{(1)} W_\mu = -\frac{1}{\mu^2}
\nabla^\nu \left( 4 {}^{({\rm E})} \! g^{\lambda\sigma}
{}^{({\rm E})} \! g^{\alpha\beta} {}^{({\rm E})} \! R_{\alpha\nu\lambda\mu}
{}^{(1)} \! g_{[\sigma \beta]} - 2
({}^{({\rm E})} \! R(\Gamma) \, {}^{(1)} \! g_{[\,]})_{\mu\nu} \right) ,
\]
so that ${}^{(1)} W_\mu$ decays at least as fast as
${}^{(1)} \! g_{[\sigma \beta]}$.
Here, $({}^{({\rm E})}\! R(\Gamma) \, {}^{(1)}\! g_{[\,]})_{\mu\nu}$
denotes terms involving
the products of the background Riemann tensor with
${}^{(1)}\!g_{[\mu\nu]}$.
Taking ${}^{({\rm E})} \! R_{\mu\nu}(\Gamma) = 0$ and setting
$\lambda = 0$, we find that for large $r$,
${}^{(1)}t^{\alpha\beta} \rightarrow 0$ and
${}^{(2)}t^{\alpha\beta} \rightarrow 0$.
More importantly, because the ${}^{(i)} t^{\alpha\beta}$
($i=1,2$) go to zero so rapidly,
the energy-momentum fluxes (see \cite{bib:Papapetrou})
\[
{}^{(i)}\! R^\mu = \int_{t_1}^{t_2} dt
\int_S {}^{(i)} t^{\mu j} n_j \, dS
\]
($i=1,2$)
vanish as the radius of the surface $S$ becomes large.
Here, the integration is carried out over a region bounded by the
hypersurfaces $t=t_1$, $\Sigma$, and $t=t_2$.
An element of the hypersurface $\Sigma$ is written
$d\Sigma_j = n_j \, dS\,dt$, where
$dS$ is an element of a two-dimensional
sphere whose radius is $r$, where $r\rightarrow \infty$ and
$n_j$ is the normal to this sphere.

We have therefore demonstrated that, for large $r$, the
flux of energy at infinity is given strictly by its general relativistic
contributions, which are known to be positive-definite.

\section{Conclusions}

The general covariance of the NGT Lagrangian density leads to a law of
energy-momentum conservation which is an immediate generalization of
the identity $\nabla_\nu T^{\mu\nu} = 0$ of GR.
The Bianchi identities also have simple generalizations in the NGT.
The nonsymmetric tensors $C_{\mu\nu}$ and $\tilde P_{\mu\nu}$ are also
found to obey identities.
However, at this time, no physical meaning is attached to these identities.

An expansion of the NGT Lagrangian density to second-order allows it to be
re-interpreted as an ``Einstein plus fields'' theory.
In this framework, the stress-energy tensor is found to be positive-definite
for large $r$.

\acknowledgements

We would like to thank M. Clayton and N.J. Cornish
for their help and for many stimulating
discussions.
This work was supported by the Natural Sciences and Engineering
Research Council of Canada.
J. L\'egar\'e would like to thank the Government of Ontario for their
support of this work.

\appendix

\section{}
\label{sec:appendix_expressions}

We give below the expressions appearing in
the calculation of the stress-energy tensor.

\begin{eqnarray*}
\delta{}^{(2)}\!g
&=& \frac{3}{4} {}^{(1)}\!g^{\rho\sigma}{}^{({\rm E})}\!g^{\beta\nu}
{}^{(1)}\!g_{\beta\nu}\delta{}^{({\rm E})}\!g_{\rho\sigma}
- {}^{(2)}\!g^{\rho\sigma}\delta{}^{({\rm E})}\!g_{\rho\sigma}
+ {}^{(1)}\!g^{\nu\sigma}{}^{({\rm E})}\!g^{\rho\beta}
{}^{(1)}\!g_{\nu\beta}\delta{}^{({\rm E})}\!g_{\rho\sigma}
\end{eqnarray*}

\begin{eqnarray*}
\delta{}^{({\rm E})}\Gamma^\sigma_{\lambda\eta}
&=& \frac{1}{2}{}^{({\rm E})}\!g^{\sigma\xi}
\left(\nabla_\eta\delta{}^{({\rm E})}\!g_{\lambda\xi}
+ \nabla_\lambda\delta{}^{({\rm E})}\!g_{\xi\eta}
- \nabla_\xi\delta{}^{({\rm E})}\!g_{\eta\lambda}\right) \\
\delta{}^{(1)}\Gamma^\sigma_{\lambda\eta}
&=& - \frac{1}{2}{}^{({\rm E})}\!g^{\sigma\alpha}
{}^{({\rm E})}\!g^{\beta\xi}
\left(\nabla_\eta {}^{(1)}\!g_{\lambda\xi}
+ \nabla_\lambda {}^{(1)}\!g_{\xi\eta}
- \nabla_\xi {}^{(1)}\!g_{\eta\lambda} \right)
\delta{}^{({\rm E})}\!g_{\beta\alpha} \\
& & \mbox{}
- \frac{1}{2}{}^{({\rm E})}\!g^{\sigma\xi}
\biggl({}^{(1)}\!g_{\alpha\xi}\delta^\beta_\lambda\delta^\gamma_\eta
+ {}^{(1)}\!g_{\lambda\alpha}\delta^\beta_\xi\delta^\gamma_\eta
+ {}^{(1)}\!g_{\alpha\eta}\delta^\beta_\xi\delta^\gamma_\lambda
+ {}^{(1)}\!g_{\xi\alpha}\delta^\beta_\eta\delta^\gamma_\lambda \\
& & \mbox{}
- {}^{(1)}\!g_{\alpha\lambda}\delta^\beta_\eta\delta^\gamma_\xi
- {}^{(1)}\!g_{\eta\alpha}\delta^\beta_\lambda\delta^\gamma_\xi
\biggr)
\delta{}^{({\rm E})}\Gamma^\alpha_{\beta\gamma} \\
\delta{}^{(2)}\Gamma^\sigma_{\lambda\eta}
&=& - \frac{1}{2}{}^{({\rm E})}\!g^{\sigma\alpha}
{}^{({\rm E})}\!g^{\beta\xi}
\left(\nabla_\eta {}^{(2)}\!g_{\lambda\xi}
+ \nabla_\lambda {}^{(2)}\!g_{\xi\eta}
- \nabla_\xi {}^{(2)}\!g_{\eta\lambda} \right)
\delta{}^{({\rm E})}\!g_{\beta\alpha} \\
& & \mbox{}
- \frac{1}{2}{}^{({\rm E})}\!g^{\sigma\xi}
\biggl({}^{(2)}\!g_{\alpha\xi}\delta^\beta_\lambda\delta^\gamma_\eta
+ {}^{(2)}\!g_{\lambda\alpha}\delta^\beta_\xi\delta^\gamma_\eta
+ {}^{(2)}\!g_{\alpha\eta}\delta^\beta_\xi\delta^\gamma_\lambda
+ {}^{(2)}\!g_{\xi\alpha}\delta^\beta_\eta\delta^\gamma_\lambda \\
& & \mbox{}
- {}^{(2)}\!g_{\alpha\lambda}\delta^\beta_\eta\delta^\gamma_\xi
- {}^{(2)}\!g_{\eta\alpha}\delta^\beta_\lambda\delta^\gamma_\xi
\biggr)
\delta{}^{({\rm E})}\Gamma^\alpha_{\beta\gamma}
- {}^{({\rm E})}\!g^{\sigma\xi}{}^{(1)}\!g^{\rho\xi}
\delta{}^{(1)}\Gamma^\rho_{\lambda\eta} \\
& & \mbox{}
+ {}^{({\rm E})}\!g^{\sigma\alpha}{}^{({\rm E})}\!g^{\xi\beta}
{}^{(1)}\!g_{\rho\xi}{}^{(1)}\Gamma^\rho_{\lambda\eta}
\delta{}^{({\rm E})}\! g_{\beta\alpha} \\
& & \mbox{}
- \frac{1}{2}\sigma\Biggl\{{}^{({\rm E})}\!g^{\sigma\alpha}
{}^{({\rm E})}\!g^{\beta\xi}
\left( {}^{(1)}g_{\xi\lambda}\delta^\mu_\eta
- {}^{(1)}g_{\eta\xi}\delta^\mu_\lambda\right)
{}^{(1)}W_\mu\delta{}^{({\rm E})}\! g_{\beta\alpha} \\
& & \mbox{}
+ \frac{1}{2}{}^{({\rm E})}\!g^{\mu\alpha}{}^{({\rm E})}\!g^{\beta\rho}
\left(\delta^\sigma_\lambda{}^{(1)}\!g_{[\rho\eta]}
+\delta^\sigma_\eta{}^{(1)}\!g_{[\rho\lambda]}\right)
{}^{(1)}W_\mu\delta{}^{({\rm E})}\!g_{\beta\alpha} \\
& & \mbox{}
- \frac{1}{2}{}^{({\rm E})}\!g^{\mu\alpha}
{}^{({\rm E})}\!g^{\beta\rho}
{}^{({\rm E})}\!g^{\sigma\xi}
{}^{({\rm E})}\!g_{\eta\lambda}
{}^{(1)}\!g_{[\rho\xi]} {}{(1)}W_\mu
\delta{}^{({\rm E})}\!g_{\beta\alpha} \\
& & \mbox{}
- \frac{1}{2}{}^{({\rm E})}\!g^{\mu\rho}
{}^{({\rm E})}\!g^{\sigma\alpha}
{}^{({\rm E})}\!g^{\beta\xi}
{}^{({\rm E})}\!g_{\eta\lambda}
{}^{(1)}\!g_{[\rho\xi]} {}{(1)}W_\mu
\delta{}^{({\rm E})}\!g_{\beta\alpha}
+ \frac{1}{2}{}^{({\rm E})}\!g^{\mu\rho}
{}^{({\rm E})}\!g^{\sigma\xi}
{}^{(1)}\!g_{[\rho\xi]} {}^{(1)}W_\mu
\delta {}^{({\rm E})}\!g_{\eta\lambda}\Biggr\}
\end{eqnarray*}

\begin{eqnarray*}
\delta{}^{({\rm E})}\!R_{\mu\nu}(\Gamma)
&=& \nabla_\beta\delta{}^{({\rm E})}\Gamma^\beta_{\mu\nu}
- \nabla_\nu\delta{}^{({\rm E})}\Gamma^\beta_{\mu\beta} \\
\delta{}^{(1)}\!R_{\mu\nu}(\Gamma)
&=& \nabla_\beta\delta{}^{(1)}\Gamma^\beta_{\mu\nu}
- \delta^\sigma_{(\nu}\delta^\rho_{\mu)}\nabla_\sigma
\delta{}^{(1)}\Gamma^\beta_{(\rho\beta)}
+ \delta^\sigma_{(\nu}\delta^\rho_{\mu)}
{}^{(1)}\Gamma^\beta_{(\alpha\beta)}
\delta{}^{({\rm E})}\Gamma^\alpha_{\rho\sigma} \\
& & \mbox{}
-\left(
{}^{(1)}\Gamma^\beta_{\alpha\nu}
\delta{}^{({\rm E})}\Gamma^\alpha_{\mu\beta}
+ {}^{(1)}\Gamma^\beta_{\mu\alpha}
\delta{}^{({\rm E})}\Gamma^\alpha_{\nu\beta}
- {}^{(1)}\Gamma^\alpha_{\mu\nu}
\delta{}^{({\rm E})}\Gamma^\beta_{\alpha\beta} \right) \\
\delta{}^{(2)}\!R_{\mu\nu}(\Gamma)
&=& \nabla_\beta\delta{}^{(2)}\Gamma^\beta_{\mu\nu}
- \delta^\sigma_{(\nu}\delta^\rho_{\mu)}\nabla_\sigma
\delta{}^{(2)}\Gamma^\beta_{(\rho\beta)}
+ \delta^\sigma_{(\nu}\delta^\rho_{\mu)}
{}^{(2)}\Gamma^\beta_{(\alpha\beta)}
\delta{}^{({\rm E})}\Gamma^\alpha_{\rho\sigma} \\
& & \mbox{}
-\left(
{}^{(2)}\Gamma^\beta_{\alpha\nu}
\delta{}^{({\rm E})}\Gamma^\alpha_{\mu\beta}
+ {}^{(1)}\Gamma^\beta_{\mu\alpha}
\delta{}^{({\rm E})}\Gamma^\alpha_{\nu\beta}
- {}^{(1)}\Gamma^\alpha_{\mu\nu}
\delta{}^{({\rm E})}\Gamma^\beta_{\alpha\beta} \right) \\
& & \mbox{} -{}^{(1)}\Gamma^\beta_{\alpha\nu}
\delta{}^{(1)}\Gamma^\alpha_{\mu\beta}
-\delta{}^{(1)}\Gamma^\beta_{\alpha\nu}
{}^{(1)}\Gamma^\alpha_{\mu\beta}
+ {}^{(1)}\Gamma^\beta_{(\alpha\beta)}
\delta{}^{(1)}\Gamma^\alpha_{\mu\nu}
+ \delta{}^{(1)}\Gamma^\beta_{(\alpha\beta)}
{}^{(1)}\Gamma^\alpha_{\mu\nu} .
\end{eqnarray*}

\end{document}